\documentclass[twocolumn,prl,aps,epsfig,floats,showpacs]{revtex4}
\def\DESepsf(#1 width #2){\epsfxsize=#2 \epsfbox{#1}}
\usepackage{epsfig}
\usepackage{graphicx}
\def\bmatrix{\left[\begin{array}}
\def\ematrix{\end{array}\right]}

\begin{document}
\title{\boldmath
$\bar B^0\to \phi K_S$ and $\bar K^{*0}\gamma$ CP Asymmetries from
 Supersymmetric Right-handed \\ Flavor Mixing:
     Implications for Heavy Quark Phenomenology  
}
\vfill
\author{Chun-Khiang Chua,\footnote{
Present address: Institute of Physics, Academia Sinica, Taipei. }
Wei-Shu Hou and Makiko Nagashima }
\affiliation{ \rm Department of Physics, National Taiwan
University, Taipei, Taiwan 10764, R.O.C. }

%
%
\vfill
\begin{abstract}
Two recent experimental developments, when combined, may have far
reaching implications.
$S_{\phi K_S} < 0$, if confirmed, would imply large $s$-$b$
mixing, a new CP phase, and right-handed dynamics. Large $\Delta
m_{B_s}$ would be likely, making the $B_s$ program at hadron
machines difficult.
Reconstruction of $B$ vertex from $K_S$ at B factories, as shown
by BaBar's first measurement of $S_{K_S\pi^0}$, makes
$S_{K_S\pi^0\gamma}$ in $\bar B\to \bar K^{*0}\gamma$ accessible.
This would be a boon for B factory upgrades.
Supersymmetric Abelian flavor symmetry, independently motivated,
can realize all of this with a light $\widetilde{sb}_1$ squark. B
factory and collider studies of flavor, CP and SUSY may not be
what we had expected.

%
%
\end{abstract}
\pacs{PACS numbers:
11.30.Hv, 
12.60.Jv, 
11.30.Er, 
13.25.Hw  
}
%
\maketitle

\pagestyle{plain}

The B factories have enjoyed great success since turning on a few
years ago, and luminosity upgrades are already being discussed.
One critical issue is the {\it physics} case, especially with
competition from LHCb and BTeV at hadronic machines. Two recent
developments strengthen this case:
the possibility that $S_{\phi K_S} < 0$, which suggests New
Physics (NP); the possibility to reconstruct the $B$ vertex from
$K_S$ at B factories, which makes $S_{K_S\pi^0\gamma}$ in $\bar
B\to \bar K^{*0}\gamma$ accessible. The former may make large
$\Delta m_{B_s}$ hard to avoid, hence dampen the prospects for CP
studies via $B_s$ system. In contrast, $S_{K_S\pi^0\gamma}$ in
$\bar B\to \bar K^{*0}\gamma$ is of great interest, as it is free
from hadronic uncertainties. Taking in view the slow start of the
Tevatron Run II program, a timely upgrade of B factories to ${\cal
L} > 10^{35}$ cm$^{-2}$s$^{-1}$ seems desirable.

In 2002, the B factories reported~\cite{Belle02,BaBar02} an
indication for ``wrong sign" mixing dependent CP asymmetry in
$\bar B^0 \to \phi K_S$, i.e. $S_{\phi K_S} = -0.39\pm0.41$.
The 2003 updates, $S_{\phi K_S} = - 0.96\pm 0.50^{+0.09}_{-0.11}$
(Belle, 140 fb$^{-1}$~\cite{Belle03}), $0.45\pm 0.43\pm 0.07$
(BaBar, 110~fb$^{-1}$~\cite{Browder}), however, are in 2.1$\sigma$
disagreement. The Belle result agrees with their previous
$-0.73\pm 0.64\pm 0.22$~\cite{Belle02} using 78~fb$^{-1}$ data,
and is by itself 3.5$\sigma$ away from the expected value of
$\sin2\phi_1(\beta) = 0.736\pm 0.049$~\cite{Browder}. The BaBar
result, however, shifted by more than $1\sigma$ from their
previous $-0.18\pm 0.51\pm 0.09$~\cite{BaBar02} based on
81~fb$^{-1}$.
Another year is needed for the issue to settle, but the 2003
average~\cite{Browder},
\begin{equation}
S_{\phi K_S} = -0.15\pm0.33,
\end{equation}
is still 2.7$\sigma$ away from 0.73. The new physics hint may well
be real.
Such a large effect would require large effective $s$-$b$ mixing
and a new CP phase.
Furthermore~\cite{Kagan,Kou}, to account for $S_{\eta^\prime K_S}
\sim \sin2\phi_1 > 0$ as well, the new interaction should be
right-handed.

In this paper we point out that a class of models with approximate
Abelian flavor symmetry~\cite{Nir} (AFS) {\it and} supersymmetry
(SUSY) provides all the necessary ingredients in a natural way.
AFS implies near maximal $s_R$-$b_R$ mixing, but has no impact
with only Standard Model (SM) dynamics.
%
With supersymmetric AFS (SAFS), however, maximal $\tilde
s_R$-$\tilde b_R$ squark mixing~\cite{CH,ACH} brings forth a
single CP phase and $s_R\tilde b_R\tilde g$ couplings.

Focusing only on the 2-3 down sector, the down quark mass matrix
normalized to $m_b$ has the elements $\hat M^{(d)}_{33} \simeq 1$,
$\hat M^{(d)}_{22} \simeq \lambda^2$, where $\lambda \cong 0.22$.
Taking analogy with $V_{cb} \simeq \lambda^2$ gives $\hat
M^{(d)}_{23} \simeq \lambda^2$ also, but $\hat M^{(d)}_{32}$ is
unknown for lack of right-handed flavor dynamics. With effective
AFS~\cite{Nir}, however, the {\it Abelian} nature implies $\hat
M^{(d)}_{23}\hat M^{(d)}_{32} \sim \hat M^{(d)}_{33}\hat
M^{(d)}_{22}$, hence $\hat M^{(d)}_{32} \sim 1 \sim \hat
M^{(d)}_{33}$. This may be the largest off-diagonal term but its
effect is hidden within SM.
In SAFS, the flavor mixing extends to $\tilde s_R$-$\tilde b_R$
squarks, which can be parametrized as
%
\begin{equation}
\widetilde M^{2(sb)}_{RR}  = \left[
\begin{array}{ll}
\widetilde m_{22}^2 &   \widetilde m_{23}^2 e^{-i\sigma} \\
\widetilde m_{23}^2 e^{i\sigma}  &  \widetilde m_{33}^2
\end{array}  \right]
  \equiv R \left[
    \begin{array}{cc}
    \widetilde m_{1}^2 & 0 \\
    0 & \widetilde m_{2}^2
    \end{array}  \right]
    R^\dagger,
%
\end{equation}
where $\widetilde m^2_{ij} \sim \widetilde m^2$, the squark mass
scale, and
\begin{eqnarray}
R = \bmatrix{cc} \cos\theta & \sin\theta \\
-\sin\theta e^{i\sigma} & \cos\theta e^{i\sigma} \ematrix,
\end{eqnarray}
diagonalizes $\widetilde M^{2(sb)}_{RR}$. There is just
one~\cite{ACH} CP phase $\sigma$, which is on equal footing with
the KM phase $\delta$ as both are rooted in the quark mass matrix.
Note that $(\widetilde M^2)_{LR} = (\widetilde M^2)_{RL}^\dagger
\sim \widetilde{m} M$ is suppressed by quark mass, while
$(\widetilde M^2)_{LL}$ is CKM suppressed.


Our interest is phenomenological rather than model building. With
strong hint for new physics CP violation in $S_{\phi K_S} <0$,
where else can effects be large? A realistic model such as SAFS
allows us to be comprehensive and make more definitive
predictions. Together with a recent demonstration~\cite{Browder}
of $B$ decay vertex finding with $K_S$, we arrive at a surprising
result: the $B_d$ system, rather than $B_s$, may be more
accessible for probing CP violation in $b\to s$ transitions
induced by $\tilde s_R$-$\tilde b_R$ mixing.

Low energy constraints are serious. Even after decoupling the $d$
flavor~\cite{ACH}, stringent kaon constraints imply that
$\widetilde m$ and $m_{\tilde g}$ to be TeV scale or higher.
We enforce 
$\widetilde m_{22}^2 \cong \widetilde m_{23}^2 \cong \widetilde
m_{32}^2 \cong \widetilde m_{33}^2 \cong \widetilde m^2 \gtrsim$
TeV. By some amount of fine-tuning~\cite{ACH}, one can have a
light ``strange-beauty" squark $\widetilde{sb}_{1}$ to enhance the
impact on $b\leftrightarrow s$ transitions.
We shall see that, $S_{\phi K_S} < 0$ requires
$\widetilde{sb}_{1}$ to be light, and the gluino to be not too
heavy.

To compute short distance coefficients, we use mass basis of
Eq.~(2) rather than mass insertions, since off-diagonal elements
are large. Hadronic amplitudes are calculated in naive
factorization, since the aim is to explore new physics effects.
Details cannot be given here, but $O^{(\prime)}_{3-6}$,
$O^{(\prime)}_{7,8}$ and $O^{(\prime)}_{9,10}$ operators arise
from the strong, electromagnetic (EM) and electroweak (EW) or $Z$
penguins, respectively, while special attention would be paid to
$O^{(\prime)}_{11,12}$, the EM and chromo-dipole penguins. The
prime indicates purely new physics effects arising from
right-handed dynamics.

For illustrations, we shall take $\widetilde m_1 \simeq 200$~GeV,
$\widetilde m$ at 1, 2~TeV
and $m_{\tilde g} = $ 0.5, 0.8~TeV. One still survives~\cite{ACH}
the $b\to s\gamma$ constraint, as shown in Fig.~1(a), where we use
${\cal B}(b\to s\gamma) = (3.14\times 10^{-4}) (\vert
c_{11}\vert^2 + \vert c_{11}^\prime \vert^2)/\vert c_{11}^{\rm
SM}\vert^2$ with $c_{11}^{\rm SM}\simeq -0.31$. For $\sigma \sim
\pi$, one has constructive $LR$ chiral enhancement effect.
Otherwise, $RR$ effect dominates and $b\to s\gamma$ is very
forgiving. We see that $b\to s\gamma$ is an effective constraint
on $LR$ mixing.

\begin{figure}[t!]
\smallskip  
\centerline{{\hskip0.05cm\epsfxsize1.6 in \epsffile{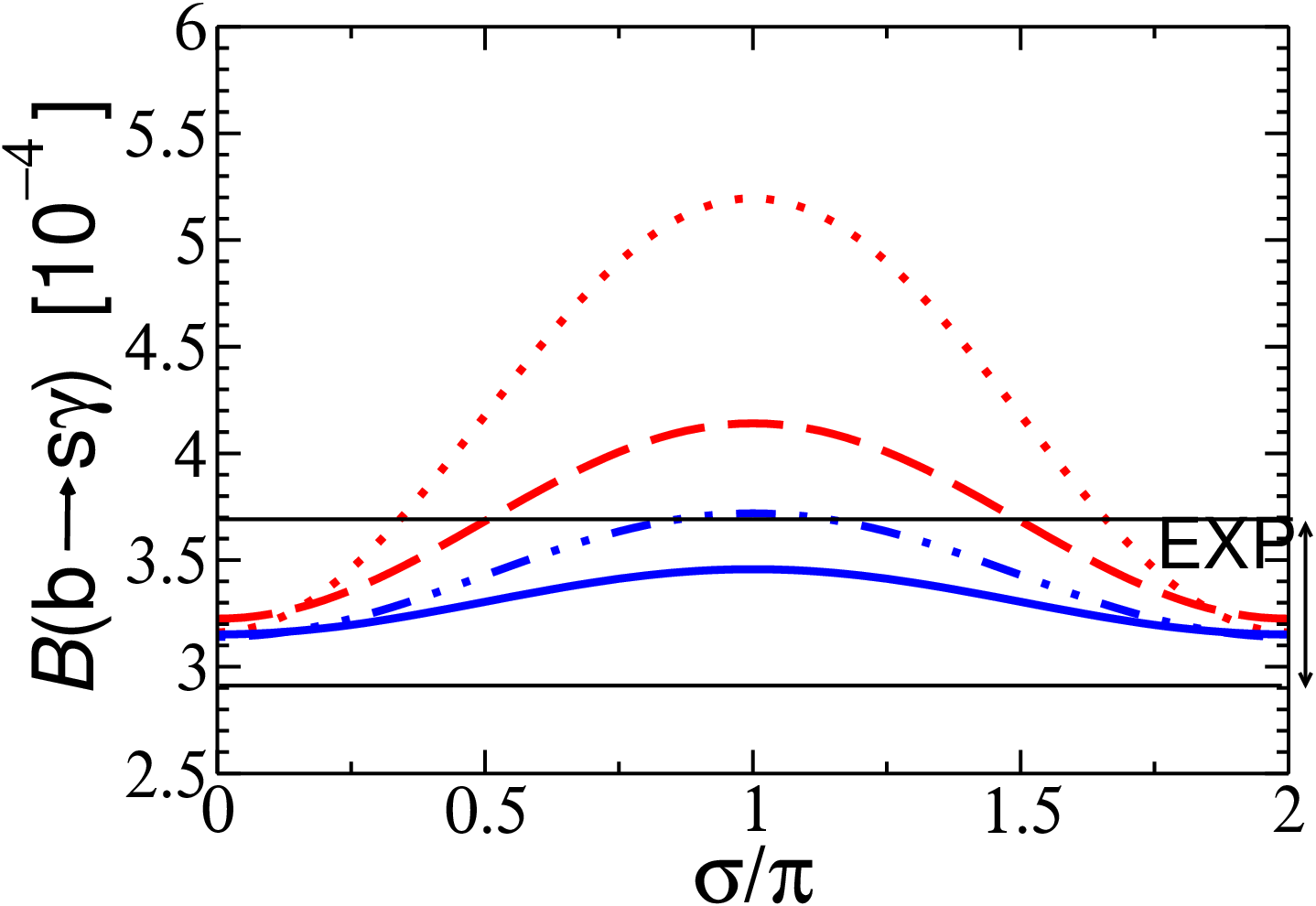}}
\hskip0.12cm
            {\epsfxsize1.59 in \epsffile{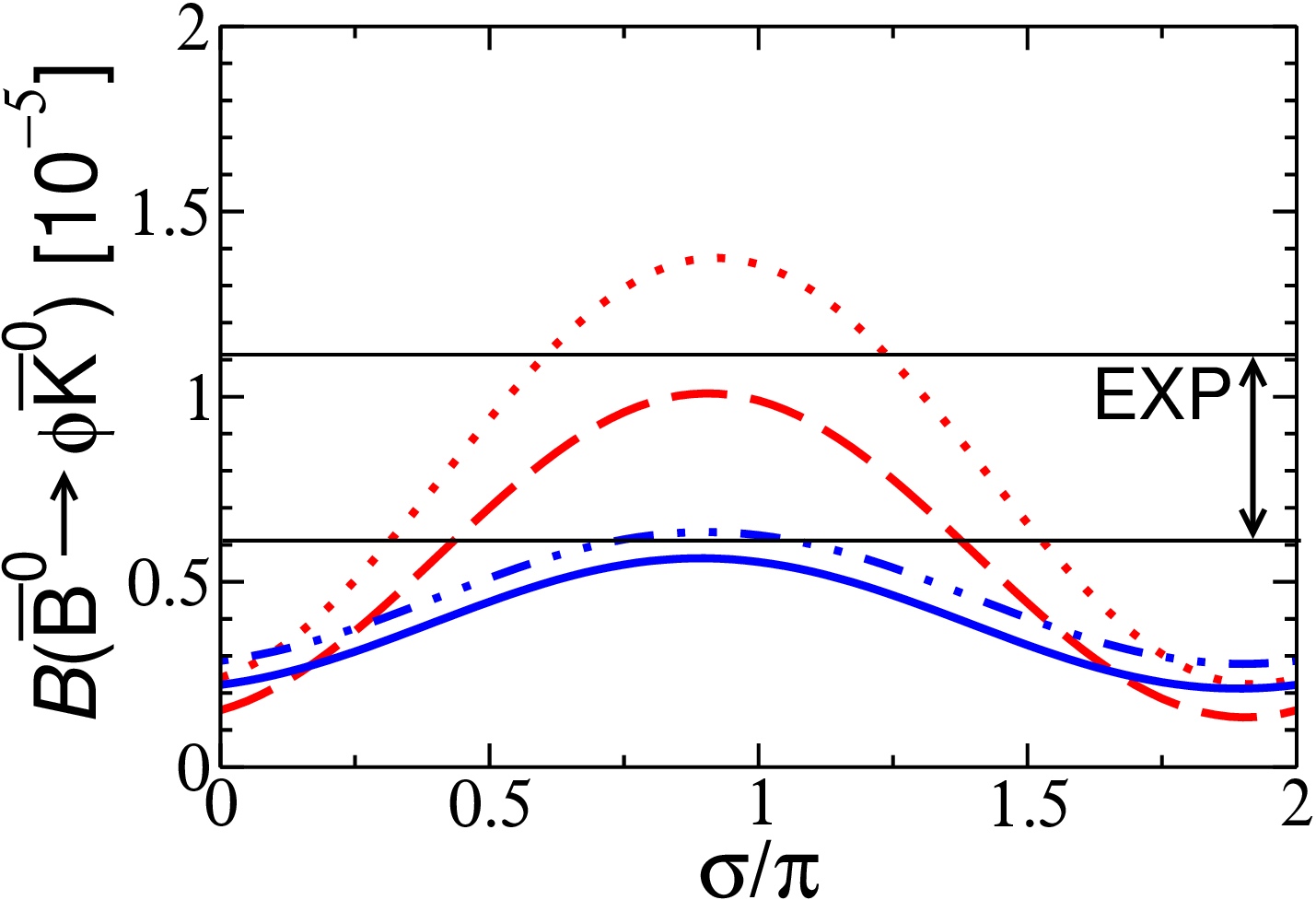}}}
\smallskip\smallskip\smallskip\smallskip\smallskip
\caption{ (a) ${\cal B}(b\to s\gamma)$ and (b) ${\cal
B}(B^0\to\phi K^0)$ vs $\sigma$ for $\widetilde m_1 = 200$ GeV
compared with experiment. Solid, dotdash (dash, dots) lines are
for $\widetilde{m} =$ 2, 1 TeV, $m_{\tilde g}=$ 0.8 (0.5) TeV. }
 \label{fig:brbsp}
\end{figure}

The $\bar B\to\phi K_S$ decay amplitude is
\begin{eqnarray}
{\cal A}(\bar B^0\to \phi \bar K^0)
 \propto \Bigl\{\cdots
 +\frac{\alpha_s}{4\pi}\frac{m_b^2}{q^2} \, \tilde{S}_{\phi K} \,
(c_{12}+c_{12}^\prime)\Bigr\},
\end{eqnarray}
where $\cdots$ are several terms $\propto a_i + a_i^\prime$, and
$q$ is the virtual gluon momentum.
It turns out that only $c_{12}^\prime$ is sensitive to the
$\widetilde{sb}_{i}$-$\tilde g$ loop. The SM strong penguin
already has a large logarithm, while the $Z$-penguin receives
large $m_t$ effect. The rate, plotted in Fig.~1(b), is compatible
with data. Comparing with Fig.~1(a), we note that for $m_{\tilde
g} = $ 0.5~TeV, the $b\to s\gamma$ and $B\to \phi K_S$ rates
balance each other at $\sigma \sim 65^\circ$, $300^\circ$. This is
supported by CP violating data, which further selects the former
branch.
%

As shown in Fig.~2(a), for $\sigma \sim 40^\circ$--$90^\circ$,
large $\tilde s_R$-$\tilde b_R$ mixing can
indeed~\cite{Kou,Murayama} turn $S_{\phi K_S}$ negative, while for
$\sigma \sim 180^\circ$--$360^\circ$, it is larger than the SM
value of $0.73$.
With $\widetilde{m}_{1}$ held fixed, there is little difference
between $\widetilde{m} =$ 1 and 2 TeV. The effect weakens for
$\widetilde{m}_{1} >$ 200~GeV, but for lighter
$\widetilde{m}_{1}$, e.g. 100~GeV, the change is not dramatic.
This is why we chose $\widetilde{m}_{1} =$ 200~GeV.
Although $m_{\tilde g} \lesssim 500$ GeV is preferred, lowering
$m_{\tilde g}$ further can lead to trouble with low energy
constraints. However, two hadronic parameters, $\tilde{S}_{\phi
K}$ and $q^2$, accompany $c_{12}^\prime$. The former arises from
evaluating the matrix element of $O_{12}^{(\prime)}$ and may be
larger than the naive factorization result of $\tilde{S}_{\phi K}
= -1.3$. For the latter, $q^2 < m_b^2/3$ is possible (within
$m_b^2/2 \gtrsim q^2 \gtrsim m_b^2/4$~\cite{GH}). Thus, $m_{\tilde
g} <$ 500~GeV may not be needed if $m_b^2\,\vert\tilde{S}_{\phi
K}\vert/q^2 > 3.9$.
Our parameter choice has been in part to reflect Eq.~(1), the
current average central value for ${S}_{\phi K}$.

\begin{figure}[t!]
\smallskip\smallskip    
\centerline{{\hskip0.05cm\epsfxsize1.55 in \epsffile{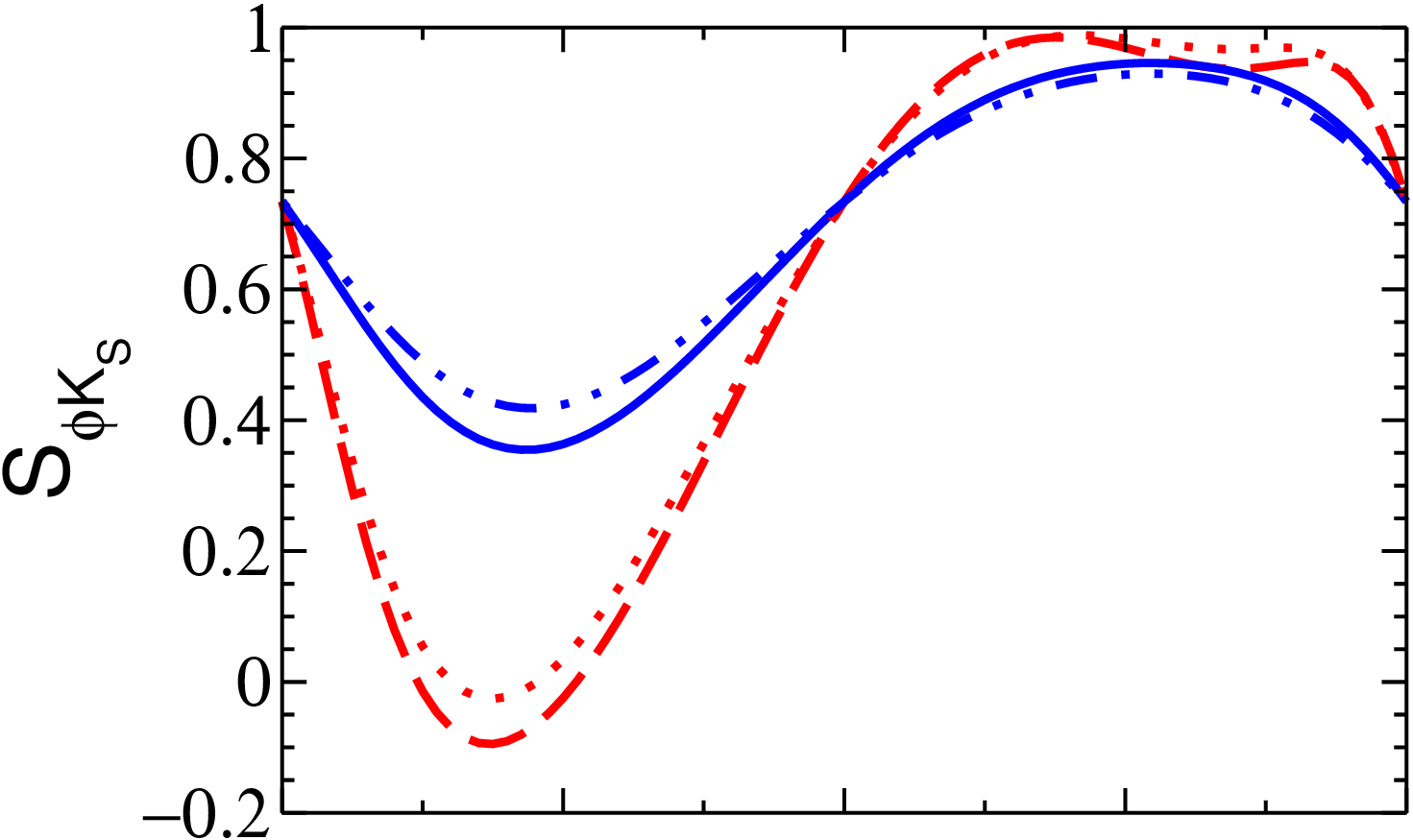}}
\hskip0.12cm
            {\epsfxsize1.55 in \epsffile{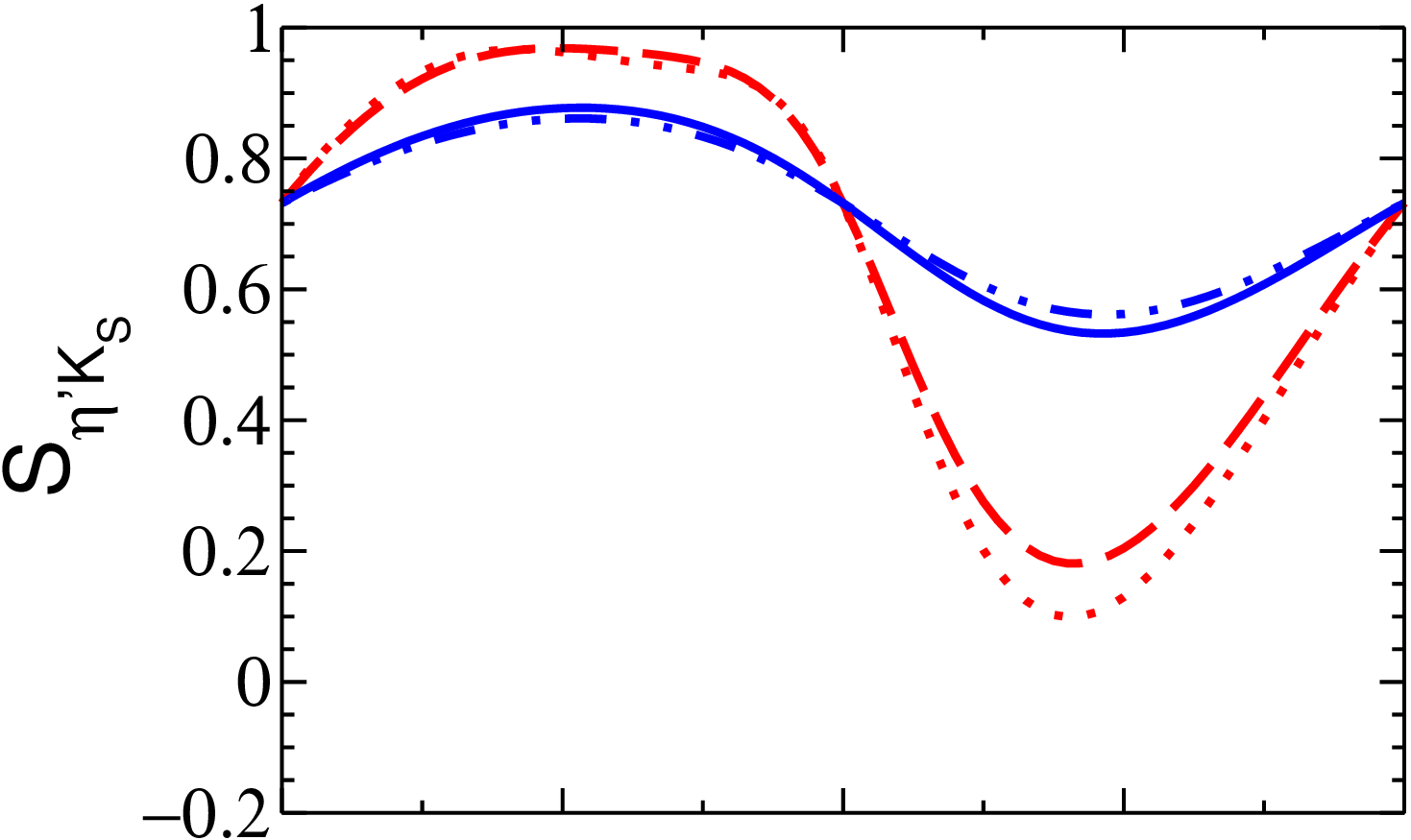}}}
\vskip0.385cm
\centerline{{\hskip0.11cm\epsfxsize1.55 in \epsffile{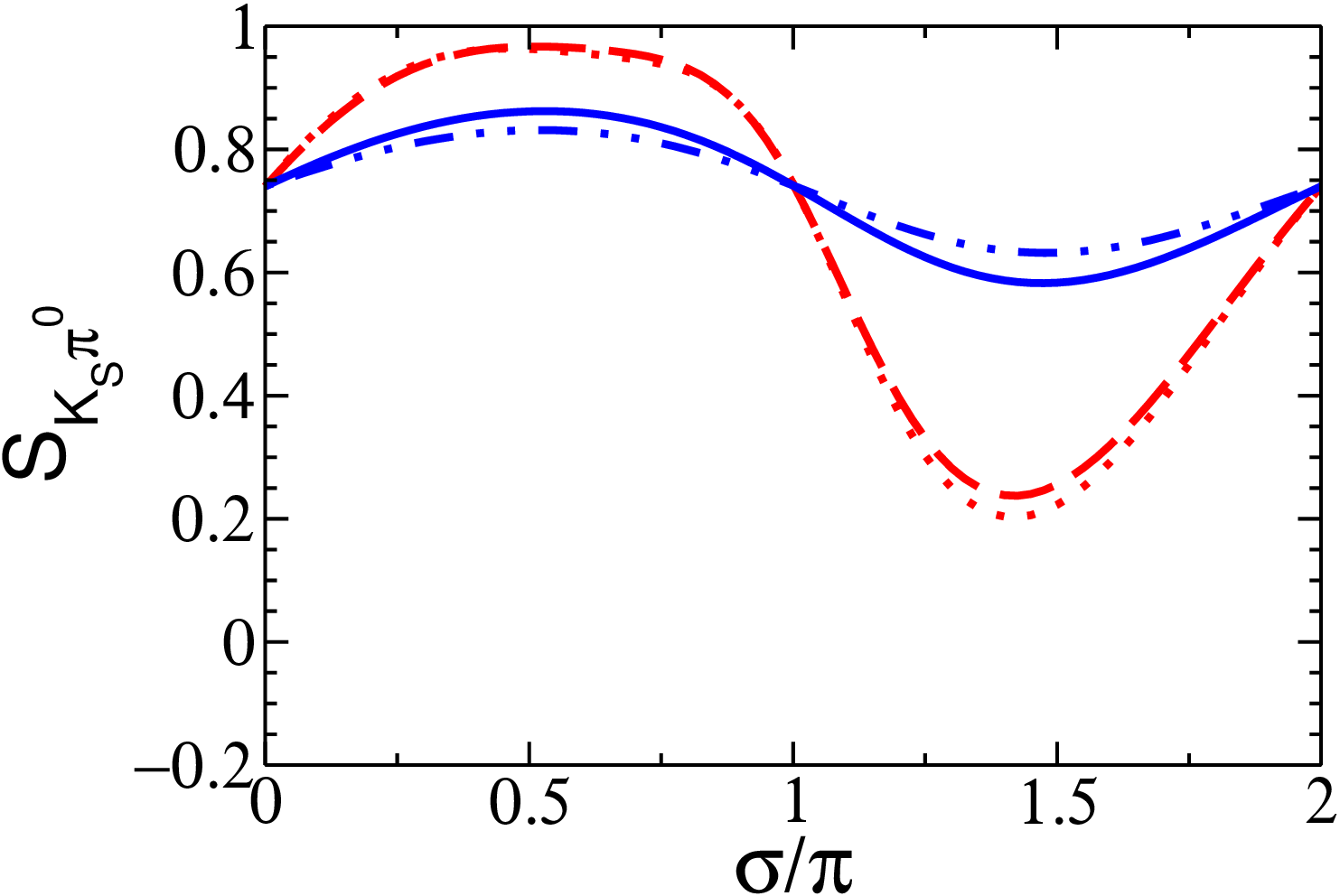}}
\hskip0.1cm
            {\epsfxsize1.55 in \epsffile{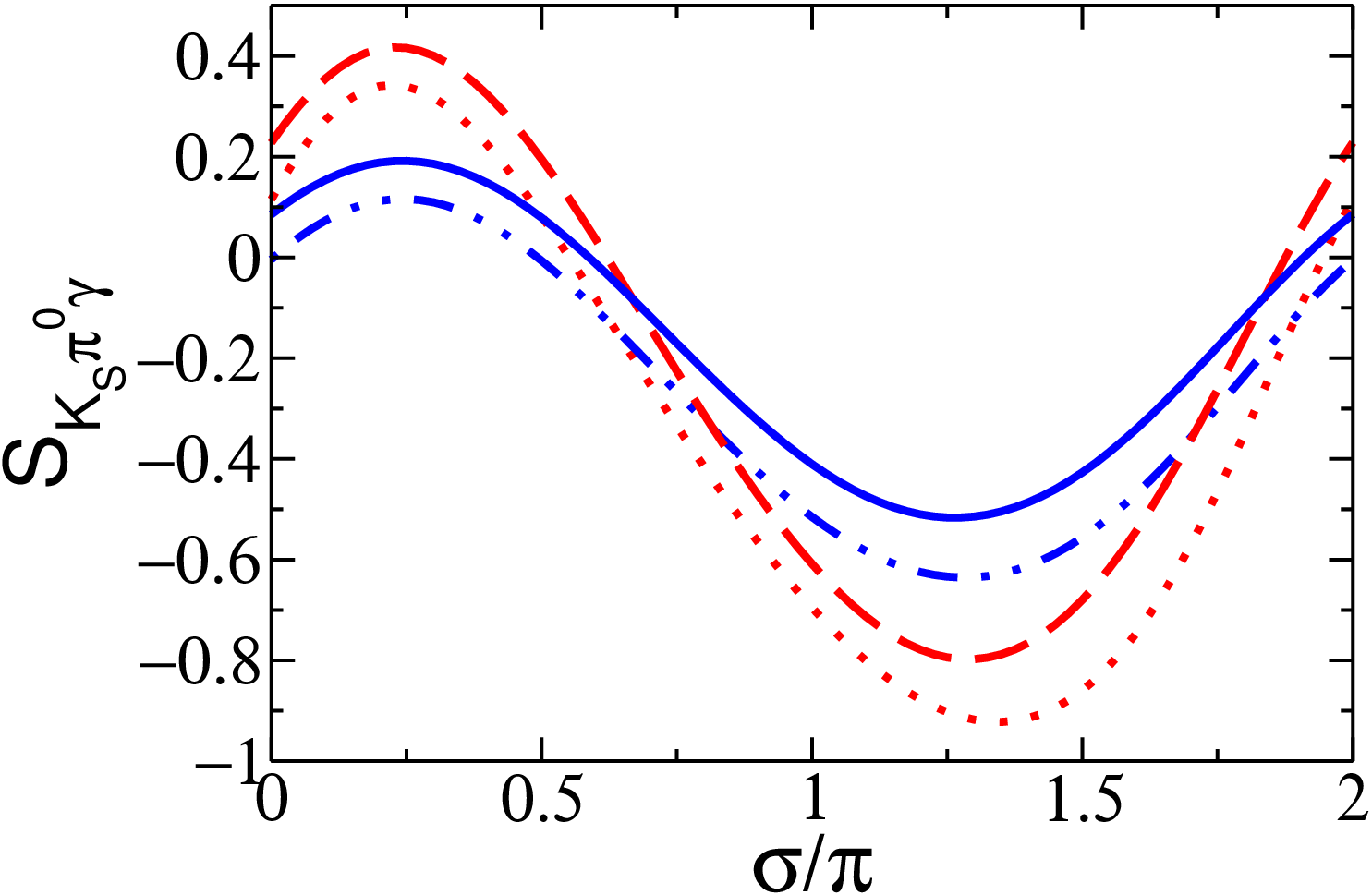}}}
\smallskip\smallskip\smallskip\smallskip
\caption{(a) $S_{\phi K_S}$, (b) $S_{\eta^\prime K_S}$, (c)
$S_{K_S\pi^0}$ and (d) $S_{K_S\pi^0\gamma}(\bar B^0\to \bar
K^{*0}\gamma)$ vs $\sigma$ with notation as in Fig. 1. }
\label{fig:Ss}
\end{figure}

Eq.~(1) is also opposite in sign with respect to~\cite{Belle03}
$\bar B\to\eta^\prime K_S$ which is also dominantly $b\to s$
penguin \cite{KKKs}. The large rate for this mode is not well
understood, which the $\widetilde{sb}_{1}$ squark does not help to
explain. However, the effect on $S_{\eta^\prime K_S}$
anticorrelates~\cite{Kou,Kagan} with $S_{\phi K_S}$. As
illustrated in Fig.~2(b), for $S_{\phi K_S} < 0$, one has
$S_{\eta^\prime K_S} \gtrsim 0.73$ which is consistent with
experiment. This is traced to the $c_{12}^\prime$ dependence in
decay amplitude,
\begin{equation}
{\cal A}(\bar B^0\to \eta^\prime  \bar K^0) \propto \Bigl\{\cdots
+\frac{\alpha_s}{4\pi}\frac{m_b^2}{q^2} \, \tilde{S}_{\eta^\prime
K} \, (c_{12}-c_{12}^\prime) \, \Bigr\}.
\end{equation}
The $\cdots$ are many terms $\propto a_i - a_i^\prime$. Even the
$c_{12}-c_{12}^\prime$ term has two terms, i.e.
$\tilde{S}^{(d)}_{\eta^\prime K} = -1.5$ and
$\tilde{S}^{(s)}_{\eta^\prime K} = -3.6$ for $B\to \eta^\prime$,
$K$ transitions, respectively. Compared to Eq.~(4), the primed
terms change sign because, in contrast to the vector current
production of $\phi$, current production of pseudoscalars
distinguishes the sign of the axial part of $V\mp A$ current.

We see that the right-handed strange-beauty squark
$\widetilde{sb}_{1}$ can generate the CP effects observed in $\bar
B\to\phi K_S$ vs $\bar B\to\eta^\prime K_S$. It is interesting
that $\sigma \sim 65^\circ$ agrees quite well with what is
inferred from Fig. 1 with rates. Note that the hadronic parameter
$\tilde{S}_{\eta^\prime K_S}$ is more involved than
$\tilde{S}_{\phi K_S}$. The extra hadronic effects needed to
enhance $B\to\eta^\prime K_S$ would also likely dilute~\cite{Kou}
the impact of the $\widetilde{sb}_{1}$-$\tilde g$ loop, unless the
latter is an active ingredient. Thus, Fig.~2(b) is only
illustrative.

Direct CP asymmetries ($A_{\rm CP})$ are very sensitive to
hadronic phases, but they are of great interest. Keeping only
perturbative penguin phases, for $(\widetilde m_1,\ \widetilde m,\
m_{\tilde g}) =$ (0.2, 1, 0.5)~TeV,
we find $A_{\phi K_S} \sim +0.21$ and $A_{\eta^\prime K_S} \sim
-0.22$. The current average~\cite{Jawahery} for the former
(latter) is slightly positive (negative), though still consistent
with zero.
More interesting is $A_{\rm CP}(K^-\pi^+)$, where the current
average~\cite{Jawahery} of $-0.09\pm 0.03$ is getting significant.
We find $A_{\rm CP}(K^-\pi^+) \simeq A_{\rm CP}(K^-\pi^0)$ drops
from $+0.12$ to $-0.02$ for $\sigma \sim 0$--$75^\circ$, which
improves upon the positive value in QCD factorization. There is
currently a disagreement between Belle and BaBar on $A_{\rm
CP}(K^-\pi^0)$, giving a zero average, but $A_{\rm CP}(\bar
K^0\pi^+) \sim -0.15$ and $A_{K_S\pi^0} \equiv A_{\rm CP}(\bar
K^0\pi^0)\sim -0.2$ for $\sigma \simeq 65^\circ$ are also not
inconsistent with data. In fact, BaBar finds~\cite{Browder}
$C_{K_S\pi^0} \equiv -A_{K_S\pi^0} = +0.40^{+0.27}_{-0.28}\pm
0.10$, giving $A_{K_S\pi^0} < 0$.

The $\widetilde{sb}_{1}$ loop clearly improves the agreement with
current experimental trend of $A_{\rm CP}$s, and the effect can be
enhanced by the hadronic parameter $\tilde S/q^2$.
It is important to stress {\it the need for right-handed
interactions}. Had the new physics been in the left-handed sector,
such that $c_{11}$ is enhanced but $c^\prime_{11}$ remains
negligible, then $A_{\rm CP}(\phi K)$ would track the sign of
$A_{\rm CP}(K\pi)$~\cite{HHY}. The same holds for $S_{\phi K_S}$
vs $S_{\eta^\prime K_S}$, $S_{K_S\pi^0}$.

The aforementioned $C_{K_S\pi^0}$ measurement is done by an
exciting new method from BaBar~\cite{Browder}. By extrapolating
the $K_S$ momentum onto the boost axis ($e^-$ direction), {\it
which is the $B$ direction}, one can determine the $B$ decay
vertex.
The $\bar B^0\to K_S\pi^0$ mode had looked formidable,
since $\pi^0\to\gamma\gamma$ leaves no track, while $K_S\to
\pi^+\pi^-$ typically decays outside the silicon vertex detector.
BaBar has clearly benefitted from a larger vertex detector with
more layers, but analysis details are not yet available.
In view of possible future development, we plot $S_{K_S\pi^0}$ vs
$\sigma$ in Fig.~2(c). The prediction of $S_{K_S\pi^0} >$ 0.73 is
similar to ${S}_{\eta^\prime K_S}$ since both have $PP$ final
states, and is consistent with $S_{K_S\pi^0} =
0.48^{+0.38}_{-0.47}\pm 0.10$ from BaBar~\cite{Browder}.

The implication of the new BaBar method goes beyond hadronic final
states.
``Wrong helicity" photons from $b\to s\gamma$ decay would indicate
new physics, which can be probed~\cite{AGS,CHH} by mixing
dependent CP violation in exclusive radiative $\bar B^0$ decays,
\begin{equation}
{S}_{M^{0}\gamma} = \frac{2\vert c_{11}c^\prime_{11}\vert}{\vert
c_{11}\vert^2 + \vert c^\prime_{11}\vert^2} \,\xi\,
\sin\left(2\phi_{B_d}-\varphi_{11}-\varphi_{11}^\prime\right),
\end{equation}
where $\xi$ is the CP of reconstructed $M^0$ final state, and
$\phi_{B_d} = \phi_1$, $\varphi_{11}^{(\prime)}$ are the CP phases
of $B_d$ mixing and $c_{11}^{(\prime)}$, respectively. Both photon
helicities must be present for $\bar B^0$ and $B^0$ decay
amplitudes to interfere. In SM which is purely left-handed,
$c_{11}^{\prime}$ hence ${S}_{M^{0}\gamma}$ vanishes with light
quark mass. Thus, ${S}_{K^{*0}\gamma}$ is a good probe of new
physics, {\it if it can be measured}. It is exciting that the
BaBar technique makes this promising via $K^{*0}\to K_S\pi^0$.
Lack of a vertex in $\bar B^0\to \bar K^{*0}\gamma$ has heretofore
prompted us to consider the rarer $\bar B^0\to \rho^0\gamma$,
$\bar K_1(1270)^0\gamma$ (both not yet seen), $\phi K_S \gamma$,
or wait for $B_s^0\to \phi\gamma$. 

Eq.~(6) shows that ${S}_{K^{*0}\gamma}$ is free from hadronic
effects that plague the hadronic modes of Fig.~2(a)--(c). The
$B\to K^*$ form factor drops out from the ratio that gives
${S}_{K^{*0}\gamma}$.
We stress that $c_{11}^{(\prime)}$, $\varphi_{11}^{(\prime)}$ and
hence ${S}_{K_S\pi^0\gamma}$ are calculable within the present
framework. For our parameter values, $\sin2\theta \equiv 2\vert
c_{11}c^\prime_{11}\vert/(\vert c_{11}\vert^2 + \vert
c^\prime_{11}\vert^2) \sim 0.15$ to 1 for $\sigma \sim 0$ to $\pi$
($2\pi$ to $\pi$), and is finite. For $\bar B^0\to \bar
K^{*0}\gamma$, we plot ${S}_{K_S\pi^0\gamma}$ vs $\sigma$ in
Fig.~2(d).
For $\sigma \sim 65^\circ$ as favored by $S_{\phi K_S}\lesssim 0$,
we see that ${S}_{K_S\pi^0\gamma}\sim$ 0.2--0.3 is expected. {\it
The measurement of this confirming effect should be pursued with
vigor at B factories}.
We remark that $b\to s\ell^+\ell^-$ rate is unaffected, since the
$Z$ penguin correction is suppressed by LR mixing. But
$c_{11}^\prime$ can be probed by the forward-backward asymmetry
for very low $m_{ee}$ when $c_{11}^{(\prime)}$ is dominant.

\begin{figure}[t!]
\smallskip\smallskip    
\centerline{{\hskip0.1cm\epsfxsize1.5 in \epsffile{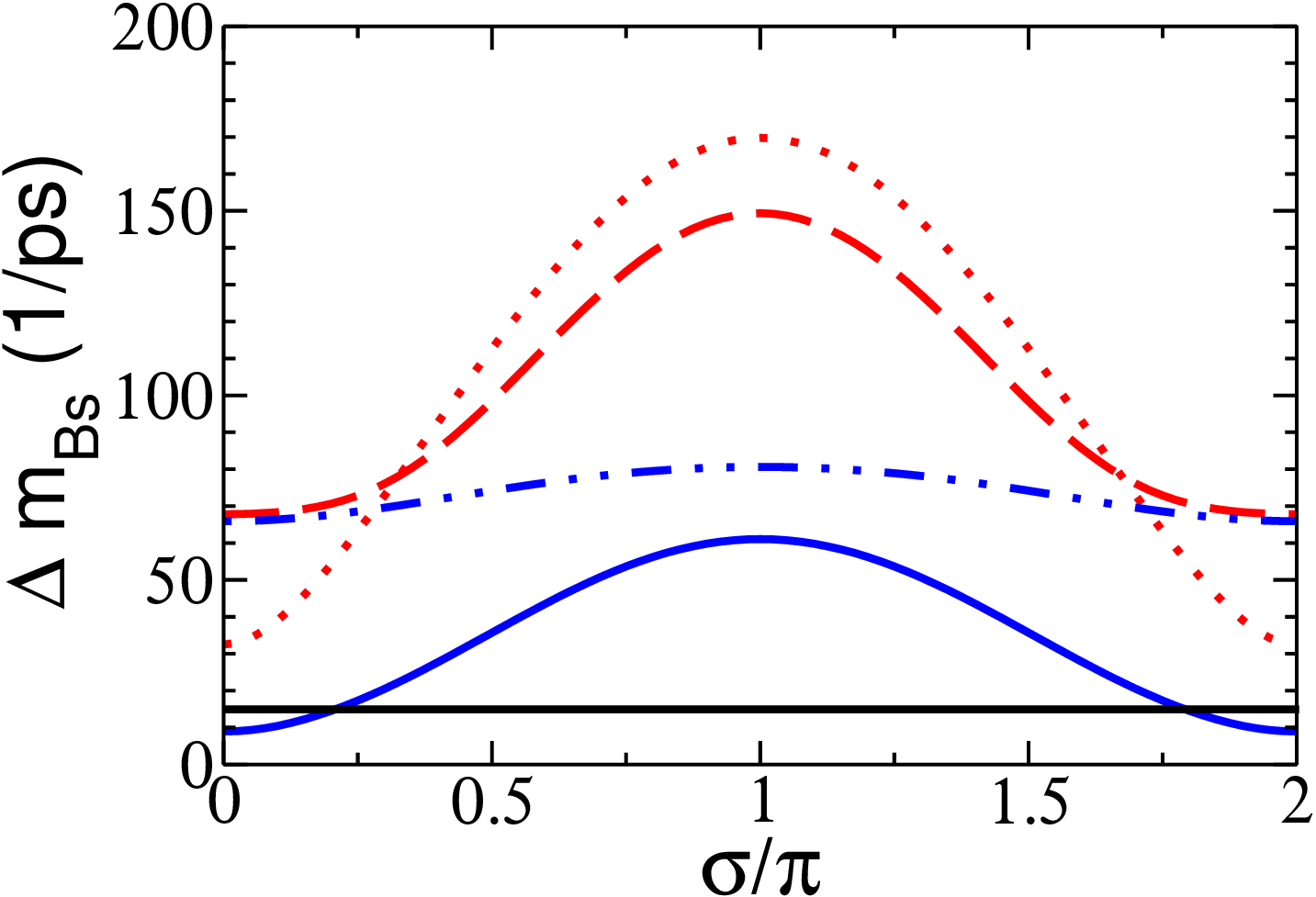}}
\hskip0.2cm
            {\epsfxsize1.5 in \epsffile{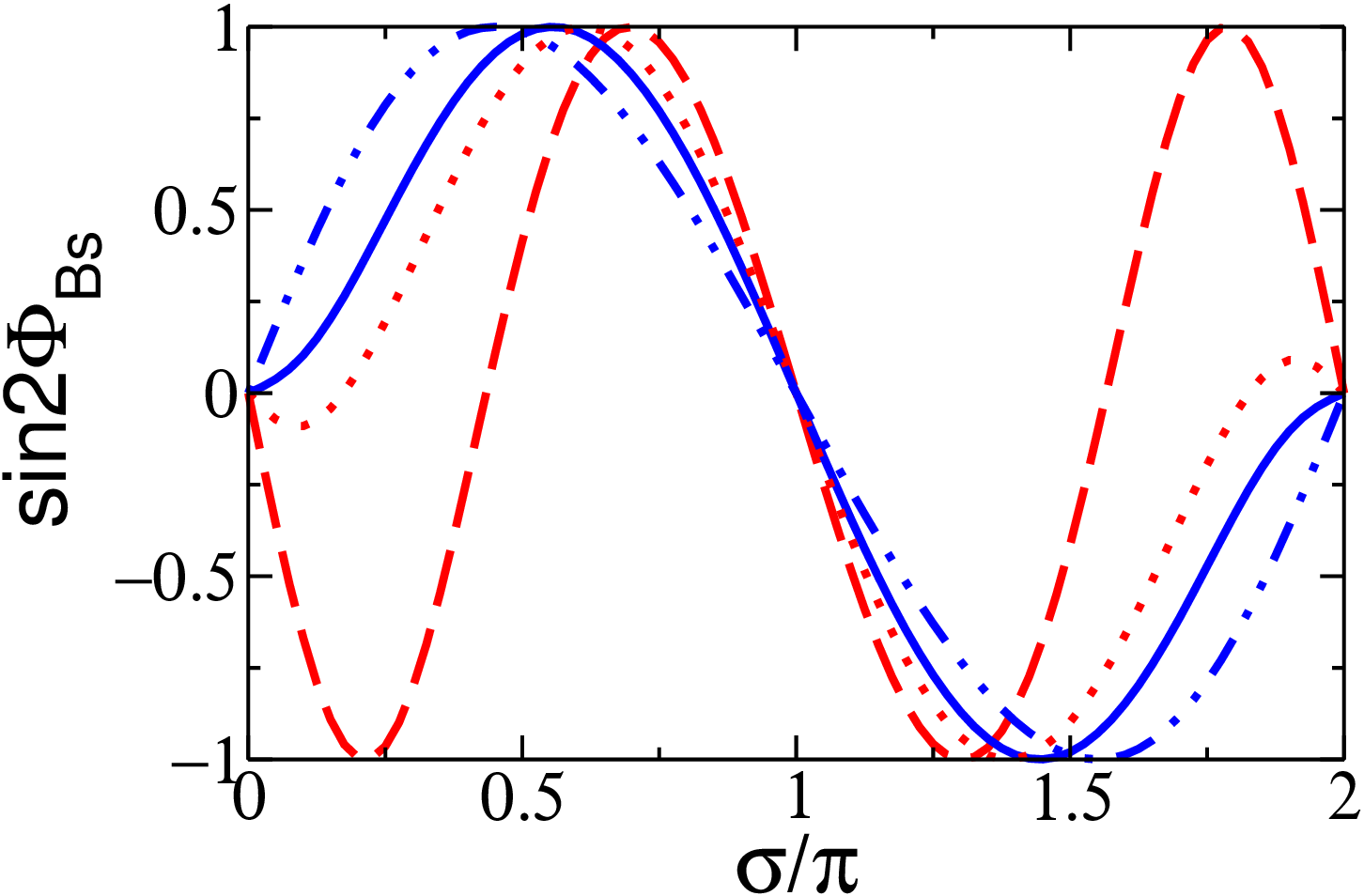}}}
\smallskip\smallskip    
\caption{ (a) $\Delta m_{B_s}$ and (b) $\sin 2\Phi_{B_s}$ vs
$\sigma$ with notation as in Fig. 1. The horizontal line is
current bound on $\Delta m_{B_s}$. }\label{fig:bsmix}
\end{figure}

Two of us have emphasized~\cite{ACH} $B_s$ mixing and CP violation
as good places to look for the effect of $\widetilde{sb}_{1}$,
with $\Delta m_{B_s}$ just above present bound of 14.9
ps$^{-1}$~\cite{PDG} as most interesting.
However, the large effect of $S_{\phi K_S} < 0$ calls for rather
light $\widetilde{sb}_1$ and $\tilde g$, and we see from Fig.~3(a)
that {\it $\Delta m_{B_s}\gtrsim$ 70 ps$^{-1}$ is hard to avoid}.
We believe this is a generic feature, not just a consequence
within SAFS. Basically, the $LR$ mixing possibility is constrained
by $b\to s\gamma$, hence $S_{\phi K_S}$ and $\Delta m_{B_s}$ are
closely linked. Measurement of $\Delta m_{B_s}\gtrsim$ 70
ps$^{-1}$ at Tevatron Run II is basically hopeless, and would be
challenging even for LHCb and BTeV. Although $\sin 2\Phi_{B_s}$
(Fig. 3(b)) can still be measured together with $\Delta m_{B_s}$,
the very fast $B_s$ oscillations would make CP studies in $B_s$
system such as $B_s\to D_S K$ very challenging.

It is intriguing that $S_{\phi K_S} <0$ implies very large $\Delta
m_{B_s}$ mixing and finite $\sin 2\Phi_{B_s}$, but they push the
limits of current vertexing technology and the CP program in $B_s$
system looks difficult. Modes such as $\bar B_s\to \phi\gamma$ for
wrong helicity photon study, and all the hadronic modes in analogy
to those of $\bar B_d$ decay, such as $\bar B_s\to
\phi\eta^\prime$, become difficult.
In contrast, $\bar B_d$ decays become more revealing as we have
illustrated, and luminosity upgrades to B factories would be
desirable.
Premium should be put on a larger silicon vertex detector. Note
that $S_{K_S\pi^0\gamma}$ measurement may be unique to B
factories, both for the precision electromagnetic calorimetry, and
because one knows the $B$ direction. It is also more important
since it is free from hadronic parameters.

The $\Lambda_b \to \Lambda\gamma$ decay is still unique to
hadronic machines. It can also probe~\cite{CHH,MR} wrong helicity
photons, since $\Lambda$ is expected to keep the polarization of
the $s$ quark. The effect is measured via the angular parameter
$\alpha_\Lambda$ ($=1$ in SM), where hadronic effects are also
absent. We find $\alpha_\Lambda \sim 0.6$ for $(\widetilde m_1,\
\widetilde m,\ m_{\tilde g}) =$ (0.2, 1, 0.5)~TeV, and drops with
$m_{\tilde g}$. 

Our minimal picture has 3 parameters: $\widetilde m_1\lesssim 200$
GeV, $m_{\tilde g}\lesssim 500$ GeV, and $\sigma\sim 65^\circ$,
with all other SUSY partners (except a dominantly bino
$\widetilde{\chi}_1^0$) $\gtrsim$ TeV scale. One still needs
collider input since $B$ decays give only indirect information.
Direct search for the light $\widetilde{sb}_{1}$ squark would be
imperative. In general~\cite{ACH}, one has to take into
consideration the presence of both $\widetilde{sb}_1 \to b
\widetilde{\chi}_1^0$ and $s \widetilde{\chi}_1^0$ which would
dilute the $b$-tagging effectiveness. The situation for the
Tevatron is not clear, but it should not be a problem for LHC. A
combined study at $B$ factories and colliders should be able to
determine the flavor and CP violating SUSY model parameters.

We stress before closing the importance of low energy constraints,
as well as the need for TeV scale SUSY in face of large flavor
violation. It was pointed out~\cite{Hisano} recently that the
$s$-quark chromoelectric dipole moment is related to the $b\to s$
color dipole by $\tilde s_L$-$\tilde b_L$ mixing insertion. We
have $\tilde s_L$-$\tilde b_L$ mixing $\sim \lambda^2$ and cannot
evade this constraint, which applies to all models. There are,
however, even more hadronic uncertainties here.
Second, a generic feature of quark-squark alignment (QSA)
models~\cite{Nir} is $D^0$-$\bar D^0$ mixing generated by
relegating $V_{us}\simeq \lambda$ to up-type quarks. Assuming
up-type squarks are also at 1--2 TeV, we typically get $x_D \sim
7\%$ (11\%--20\%) for $m_{\tilde g} = $ 800 (500)~GeV, which
should be compared with the bound of $x_D \equiv \Delta
m_D/\Gamma_D \lesssim 2.9\%$~\cite{PDG}. This makes a 500~GeV
gluino problematic,
%
and is a reminder of flavor or family interrelations in a
realistic setting. Work on $D^0$ mixing in QSA models should be
refined, and $D^0$ mixing should clearly be searched for.

Finally, maximal $\nu_\mu$-$\nu_\tau$ mixing
may~\cite{CMM} be related to right-handed $\tilde s_R$-$\tilde
b_R$ mixing in SUSY GUT framework. While this adds to the
attraction of a light strange-beauty squark, it involves extra
dynamical assumptions at very high scale (including right-handed
neutrino mass). Our working scale has been TeV and below, and
based on observed flavor patterns.
%

In summary,
$S_{\phi K_S} < 0$, if confirmed, would require large $s$-$b$
mixing with new CP phase and right-handed dynamics. $B_s$ mixing
would likely be large and the $B_s$ program becomes difficult. On
the other hand, $S_{K_S\pi^0\gamma} \neq 0$ in $\bar B^0\to \bar
K^{*0}\gamma$ is likely, and one now has good prospect for
measurement at B factories. A three parameter minimal model can
generate all these effects. Although SUSY is above TeV scale, the
model has 1) a right-handed, flavor-mixed ``strange-beauty" squark
$\widetilde{sb}_1$ below 200 GeV, 2) one new CP phase, and 3)
$m_{\tilde g} \lesssim 500$ GeV. It has independent motivation
from quark mass and mixing patterns with an underlying effective
Abelian flavor symmetry. Luminosity upgrades of existing B
factories would be very worthwhile, and a combined study with
colliders can determine model parameters.

\vskip 0.3cm \noindent{\bf Acknowledgement}.\ \ This work is
supported in part by grants NSC-92-2112-M-002-024,
NSC-92-2811-M-001-054 and NSC92-2811-M-002-033, the BCP Topical
Program of NCTS, and the MOE CosPA project.
WSH enjoyed discussions at the 5th Workshop on Higher Luminosity B
Factory, Izu, Japan, September 2003.


\begin{thebibliography}{99}
\bibitem{KKKs}
Since three body production is not well understood, we do not
comment on $\bar B^0 \to K^+K^-K_S$~\cite{Belle03}, which has
$\phi K_S$ cut out and is dominantly CP even.
\end{thebibliography}
\end{document}